\documentclass[prl,a4paper,reprint,showpacs]{revtex4-1}
\usepackage{color}
\usepackage{graphicx}                          
\usepackage{amssymb,amsfonts,amsmath}
\usepackage[utf8]{inputenc}
\usepackage{bbm}
\usepackage{mwe}
\usepackage[export]{adjustbox}
\usepackage{todonotes}

\begin{document}

\title{Correlations between thresholds and degrees:
An analytic approach to model attacks and failure cascades
}

\author{Rebekka Burkholz} 
\email{rburkholz@ethz.ch}
\affiliation{ETH Zurich, Institute of Machine Learning \\Universitätstrasse 6, 8092 Zurich, Switzerland}
\author{Frank Schweitzer}
\email{fschweitzer@ethz.ch}
\affiliation{ETH Zurich, Chair of Systems Design \\Weinbergstrasse 56/58, 8092 Zurich, Switzerland}
\begin{abstract}
Two node variables determine the evolution of cascades in random networks: a node's degree and threshold.
Correlations between both fundamentally change the robustness of a network, yet, they are disregarded in standard analytic methods as local tree or heterogeneous mean field approximations because of the bad tractability of order statistics. 
We show how they become tractable in the thermodynamic limit of infinite network size. 
This enables the analytic description of node attacks that are characterized by threshold allocations based on node degree.  
Using two examples, we discuss possible implications of irregular phase transitions and different speeds of cascade evolution for the control of cascades. 
\end{abstract}
\pacs{89.75.-k, 89.65.-s, 02.50.-r}
\maketitle

How robust is a complex network against attacks, for instance the targeted or random removal of nodes or links? 
This is of fundamental interest for network science  \cite{albert2000error,Barrat2008,Newman2010} and tightly linked to the study of binary state dynamics in Physics, 
for instance, in form of percolation models \cite{Saberi20151}, the zero-temperature random field Ising model \cite{IsingCascade}, fiber bundle models \cite{RevModPhys.82.499}, or models of epidemic spreading \cite{RevModPhys.87.925}. 
Similar approaches also describe phenomena diverse as opinion formation \cite{Watts2002,PhysRevLett.101.018701}, financial systemic risk \cite{ReviewFinanceContagion,Amini2012,Burkholz2015,BurkholzMultiplex}, black-out cascades in power grids \cite{PhysRevLett.118.048301}, and the resilience of food webs in ecology \cite{Polis2000473,Barrat2008}. 

Most of these models can be mapped to a set-up \cite{Lorenz2009,GleesonPairApproximation} where the binary state $s_i \in \{0,1\}$ of a node $i \in V$ is determined by two variables: its load $\lambda_i$, which  indicates the fragility of the node, and its threshold $\theta_i$, which indicates its robustness. 
A node fails ($s_i = 1$) whenever its load exceeds its threshold, $\theta_i \geq \lambda_i(t)$.
The load $\lambda_i(t)$ can change over time, measured in discrete time steps $t=0,\ldots,T$, dependent on the interaction between nodes. 
For example, in case of a failure cascade, the load of failing nodes can be redistributed to neighboring nodes. 

Consequently, one way to prevent such failure cascades is to allocate specific thresholds to nodes.
While, in general, this allocation of thresholds could be dynamic, 
in this letter, we focus on quenched networks where the thresholds and the network topology stay constant over time. 
The (fixed) degree of a node is defined by the number of its neighbors, i.e. by the links specified in the link set $E$ of a network. 
In the following, we concentrate on allocation schemes where thresholds and node degrees \emph{are correlated}.
Such correlations are empirically grounded, e.g. well connected banks have the tendency of lower equity than less connected banks \cite{Arinaminpathy2012,ReviewFinanceContagion}, even though the opposite is expected to lead to more robust systems \cite{Roukny2013,Gai2011}. 
Some models already assume that a threshold depends explicitly on the node degree \cite{Bak1987,GleesonPairApproximation,2016arXiv161208479L} or on degree-related centrality measures.
Yet, it is important to note that, dependent on the specific network, attack strategies based on centrality measures can be more effective than attacks based on node degrees \cite{Holme}.

In the following, we consider random graphs with prescribed degree distribution $p(k)$ \cite{Molloy1995,Newman.Strogatz.ea2001Randomgraphswith}.
This has the advantage that ensemble averages of interesting quantities can be derived by analytic iterative approaches, so called local tree approximations or heterogeneous mean field approximations. 
They become exact in the thermodynamic limit of infinite network size $N \rightarrow \infty$ (where $N$ is the number of nodes in the network). 
These approximations have been extended to capture degree-degree correlations $p(k,d)$. 
Yet, surprisingly, the equally fundamental correlations between degrees and thresholds have not been studied analytically. 
Only one specific case, i.e. the removal of nodes with higher degrees, has been studied for percolation \cite{Callaway2000,Cohen2001}. 
Other investigations of deterministic attacks so far rely on simulations \cite{2016arXiv161208479L,Roukny2013}, because ranking of nodes leads to order statistics that are difficult to capture analytically.
In this letter, we show that for infinitely large networks the ranks simplify significantly and we can identify them by a transformation of the respective cumulative distribution function (cdf).

Let's assume that $N$ degrees are drawn independently from a degree distribution $p(k)$ with cdf $F(k)$.
We order them so that $k_{1/N} \leq k_{2/N} \leq \ldots \leq k_{N/N}$.
Correspondingly, $N$ thresholds are drawn independently from an arbitrary threshold distribution with cdf $\Phi(\theta)$ and ordered $\theta_{1/N} \leq \theta_{2/N} \leq \ldots \leq \theta_{N/N}$.
Each node in the network receives exactly one degree and one threshold. 
A specific attack strategy $a$ defines this assignment. 
It thus decides which nodes are prone to failure because of a small threshold. 
Formally, $a$ is a bijective (Borel-measurable) function ${a:}$ $[0,1] \rightarrow [0,1]$ that assigns each degree rank $i/N$ a threshold rank $a(i/N)$.
Thus, a node has degree $k_{i/N}$ and threshold $\theta_{a(i/N)}$.
While $a$ be non-linear in general, the two most common heuristic attack strategies correspond to linear $a$. 
In case of the identity $a(x) = x$, thresholds and degrees are perfectly positively correlated. 
Nodes with higher degree receive higher thresholds and are less prone to failure, while peripheral nodes with small degree are conjectured to fail. 
We call this attack strategy the \emph{peripheral failures} (pf) scheme.
The choice $a(x) = 1-x$, on the other hand, means that degrees and thresholds are perfectly anti-correlated. 
Nodes with higher degrees receive lower thresholds and therefore are more prone to fail. 
Hence, this attack strategy is focused on \emph{hubs} and we refer to it as the \emph{central failures} (cf) scheme. 

\begin{figure}[h]
 \centering
   \includegraphics[width=0.47\textwidth]{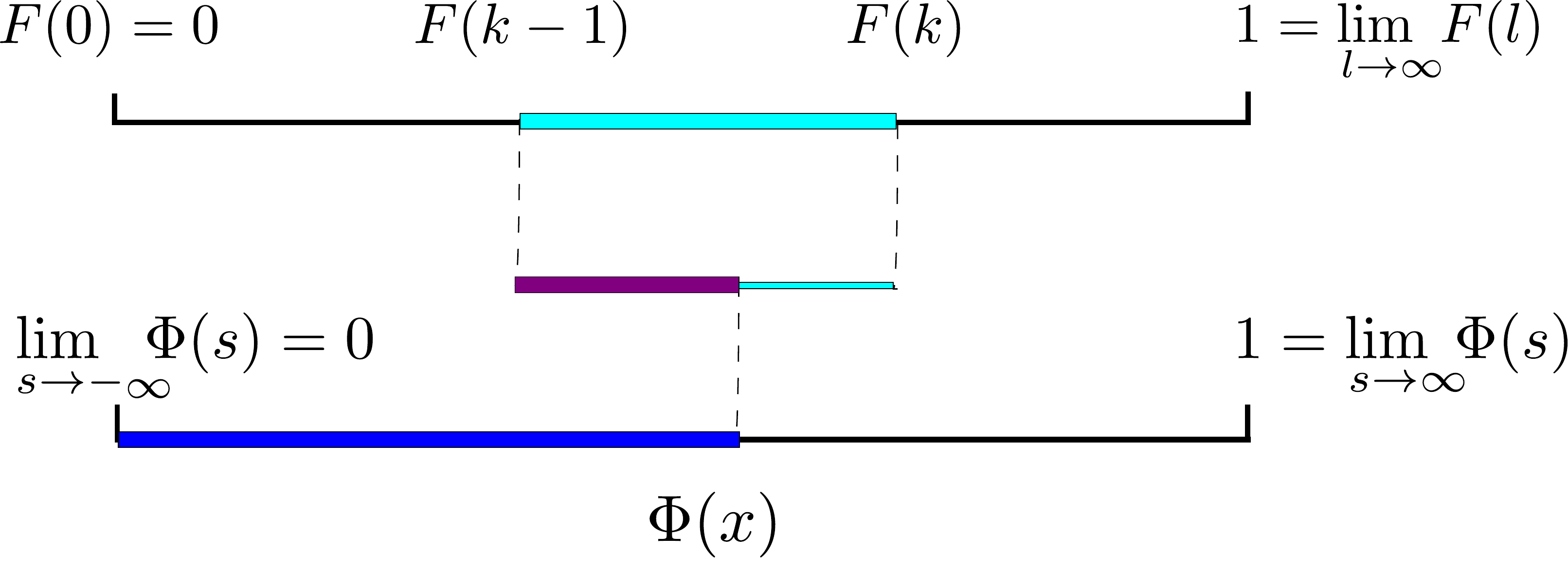}
 \caption{Illustration of the threshold assignment for central failures. The interval ${]F(k-1), F(k)]}$ (in cyan color) represents (the fraction of) all nodes with degree $k$ in the network, while the blue interval corresponds to nodes with threshold $\theta \leq x$. Their intersection, the purple interval, can thus be associated with all nodes in the network which have degree $k$ and a threshold smaller than or equal to $x$. }
\label{fig:IllustPF}
\end{figure}

As $i/N$ counts the fraction of nodes in the network with degree $k \leq k_{i/N}$, it coincides with the value of the empirical cdf $i/N = F^{\mathrm{emp}}(k_{i/N})$.  
Thus, for $N\rightarrow \infty$, a rank $i/N$ converges to a value of the theoretical cdf $F(x)$ ($i/N \rightarrow F(x)$) and, correspondingly, its threshold rank to $a(F(x))$, which belongs to a threshold $\theta$ with $a(F(x)) = \Phi(\theta)$.
Hence, for $F(x) \in ]F(k-1),F(k)]$, a node equipped with the rank $F(x)$ has degree $k$ and threshold $\Phi^{-1}\left(a(F(x))\right)$, where $\Phi^{-1}$ denotes the generalized inverse or quantile function of $\Phi$.  
Accordingly, we can express the threshold cdf $F_{\Theta(k)}\left(x\right)$ of a node conditional on its degree $k$ as 
\begin{align}\label{eq:cdf}
 F_{\Theta(k)}\left(x\right) = \frac{ \Phi \arrowvert_{\Phi^{-1}\left(a\left(]F(k-1),F(k)]\right)\right)} }{p(k)},
\end{align}
where $\Phi\arrowvert_{\mathcal{M}}$ denotes the restriction of $\Phi$ to a set $\mathcal{M}$, i.e. $\Phi\arrowvert_{\mathcal{M}}(x) = \Phi(x)$ for $x\in \mathcal{M}$ and $\Phi\arrowvert_{\mathcal{M}}(x) = 0$ for $x\notin \mathcal{M}$. 
The initial threshold cdf $\Phi$ is restricted to thresholds that belong to nodes that we can identify with the interval $]F(k-1),F(k)]$, i.e., the nodes with degree $k$. 
It is renormalized by $p(k)$, the probability mass of the respective set of nodes.
Based on this consideration, we can also compute the Spearman rank correlation coefficient between thresholds and degrees as
\begin{align}
 r_s = \sqrt{12} \frac{\int^1_0 \left(x-1/2\right) \left(a(x)- \bar{a}\right) \; dx}{\sqrt{\int^1_0 \left(a(x)- \bar{a}\right)^2 \; dx}},
\label{eq:1}
\end{align}
where $\bar{a} = \int^1_0 a(x)\; dx$ is defined as average of $a$ with respect to a uniform distribution. 
As we would expect, $r_s = 1$ for peripheral failures and $r_s = -1$ for central failures. 
Also the derivation of the conditional threshold distributions $F_{\Theta(k)}\left(x\right)$ becomes more intuitive for these two extreme cases. 

\begin{figure}[b]
 \centering
   \includegraphics[width=90mm]{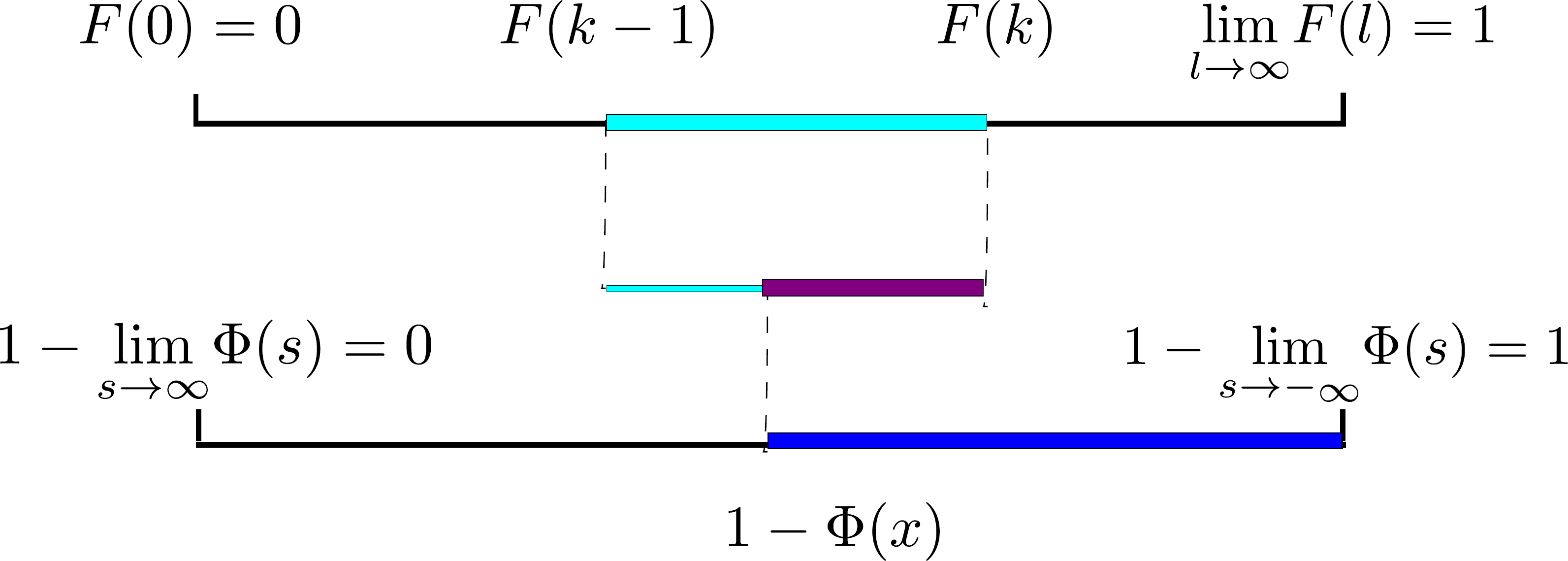}
 \caption{Illustration of the threshold assignment for central failures analogously to Fig.~\ref{fig:IllustPF}.}
\label{fig:IllustCF}
\end{figure}

In Figs.~\ref{fig:IllustPF} and \ref{fig:IllustCF}, all nodes of an infinitely large network are mapped to the interval ${[0,1]}$, where their position is defined by the degree cdf $F$.
Thus, all nodes with degree $k$ can be associated with the interval ${]F(k-1), F(k)]}$ (cyan color in Figs.~\ref{fig:IllustPF} and \ref{fig:IllustCF}).
This way, we do not depict nodes, but fractions of nodes in the network or their probability mass when randomly sampling from all nodes in the network.
Analogously, the bottom interval (in Fig.~\ref{fig:IllustPF}) corresponds to the ordered thresholds. 
For peripheral failures (Fig.~\ref{fig:IllustPF}), nodes with degree $k$ are equipped with threshold values in the interval ${]\Phi^{-1}\left(F(k)\right), \Phi^{-1}\left(F(k-1)\right)]}$.
For central failures, we have ${]\Phi^{-1}\left(1-F(k-1)\right), \Phi^{-1}\left(1-F(k)\right)]}$.
Consequently, Eq.~(\ref{eq:cdf}) simplifies for peripheral failures to
\begin{align}
\begin{split}\label{eq:leavesgeneralzero}
 & F_{\Theta^{\rm{(pf)}}(k)}(x) 
  = \frac{\left| \; ]F(k-1), F(k)] \cap ]0, \Phi\left(x\right)] \;\right|}{\left| \; ]F(k-1), F(k)] \; \right|} \\
 & = \frac{\min\left\lbrace F(k), \Phi\left(x\right)\right\rbrace  -  F(k-1)}{p(k)}\mathbbm{1}_{\left\lbrace \Phi\left(x\right) > F(k-1)\right\rbrace}(k),
\end{split}
\end{align}
where $\mathbbm{1}$ denotes the indicator function that is defined for any set $\mathcal{M}$ as $\mathbbm{1}_{\mathcal{M}}(x) = 1$ if $x \in \mathcal{M}$ and $\mathbbm{1}_{\mathcal{M}}(x) = 0$ for $x \notin \mathcal{M}$.
Formally, this formula measures the overlap between the intervals ${]F(k-1), F(k)] }$ and ${]0, \Phi\left(x\right)]}$, divided by the width $\left| \; ]F(k-1), F(k)] \; \right| = p(k)$ to normalize the cdf.
The same idea applies also to central failures, 
however, in this case the fraction of nodes are ordered decreasingly.
As illustrated in Fig.~\ref{fig:IllustCF}, the position of nodes in the interval corresponds to the probability mass $1-\Phi(x)$.
Thus, nodes with a threshold bigger or equal to $x$ correspond to the interval $[0, 1-\Phi(x)]$, while nodes with threshold smaller or equal to $x$ belong to $[1-\Phi(x), 1]$.
To calculate the fraction of failed nodes with a threshold smaller or equal to $x$ within the fraction of nodes with degree $k$, we have to estimate the overlap between the intervals $[1-\Phi(x), 1]$ and  ${]F(k-1), F(k)] }$:
\begin{align}
\begin{split}\label{eq:hubsgeneralzero}
 & F_{\Theta^{\rm{(cf)}}(k)}(x)
  = \frac{\left| \; ]F(k-1), F(k)] \cap ]1-\Phi\left(x\right), 1] \;\right|}{\left| \; ]F(k-1), F(k)] \; \right|} \\
 & = \frac{F(k)  - \max\left\lbrace F(k-1), 1-\Phi\left(x\right)\right\rbrace}{p(k)}\mathbbm{1}_{\left\lbrace F(k) > 1-\Phi\left(x\right)\right\rbrace}(k).
\end{split}
 \end{align}

To know the threshold distributions is necessary to compute average quantities in infinitely large random networks. 
The cascade size, i.e. the fraction of failed nodes $\rho(t) = 1/N\sum_i s_i(t)$, is 
of particular interest as a measure of network robustness or systemic risk.
We can express it as $\rho(t) = \sum_k p(k) \int p_{\Lambda(t)}(\lambda) F_{\Theta(k)}(\lambda) \; d\lambda$ for $N \rightarrow \infty$ \cite{Burkholz2017}. 
This form simplifies for models where the load only depends on the number of failed neighbors $m$ of a node. 
The term $F_{\Theta(k)}(\lambda)$ then refers to the response function $F_{k,m}$ in Ref. \cite{GleesonPairApproximation}.
In general, the iterative update of the load distribution $p_{\Lambda(t)}$ in time follows from the specific cascade model, for instance as in \cite{Gleeson2007,Burkholz2015,Burkholz2017}, and can also depend on multiplex network structures \cite{BurkholzMultiplex}. 
To elucidate our approach, we use two well studied and generic models that have been termed \emph{exposure diversification} (ED) and \emph{damage diversification} (DD).
The ED model, introduced in \cite{Watts2002}, was applied to different fields, including opinion formation \cite{Watts441} and Finance \cite{Battiston2012a,Roukny2013}. 
It is based on the idea that a node simply carries the fraction of its failed neighbors as load.
Thus, the failure of hubs usually has devastating consequences, as it impacts many nodes. 
The cascade size can be reduced by protecting nodes with higher degree, by assigning them higher thresholds (pf)  without changing the overall threshold distribution.
The DD model is a cascade model variant \cite{Lorenz2009}, where each failing node $j$ spreads the load $1/k_j$ to each of its neighbors.
Hence, the load that single neighbors receive from a failing hub is rather small. 
Thus, the negative impact of failing hubs is counteracted.
The cascade size can be reduced by protecting nodes with smaller degree, i.e by assigning them higher thresholds (cf).

\begin{figure}[t]
 \centering
\includegraphics[width=0.238\textwidth]{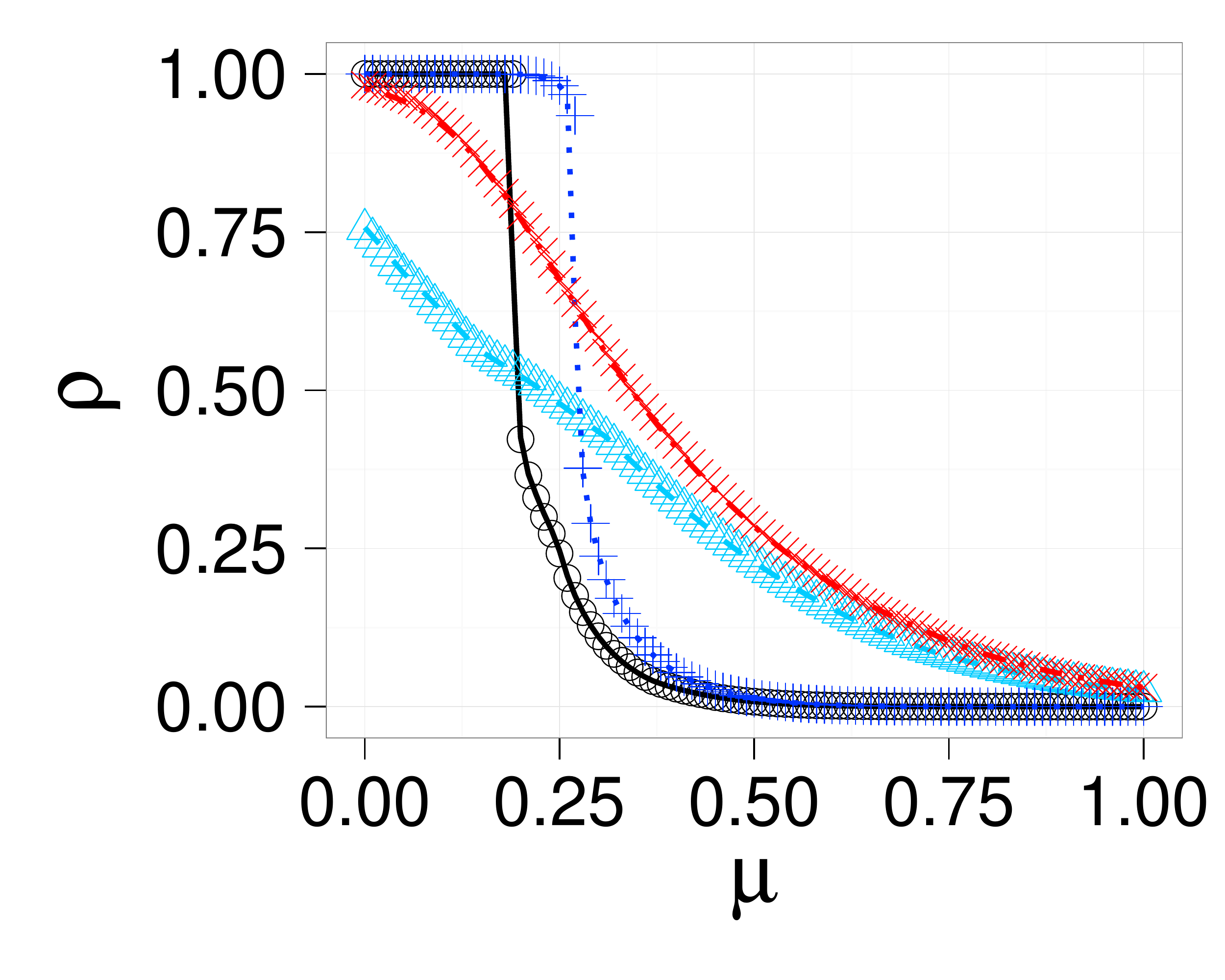}
\includegraphics[width=0.238\textwidth]{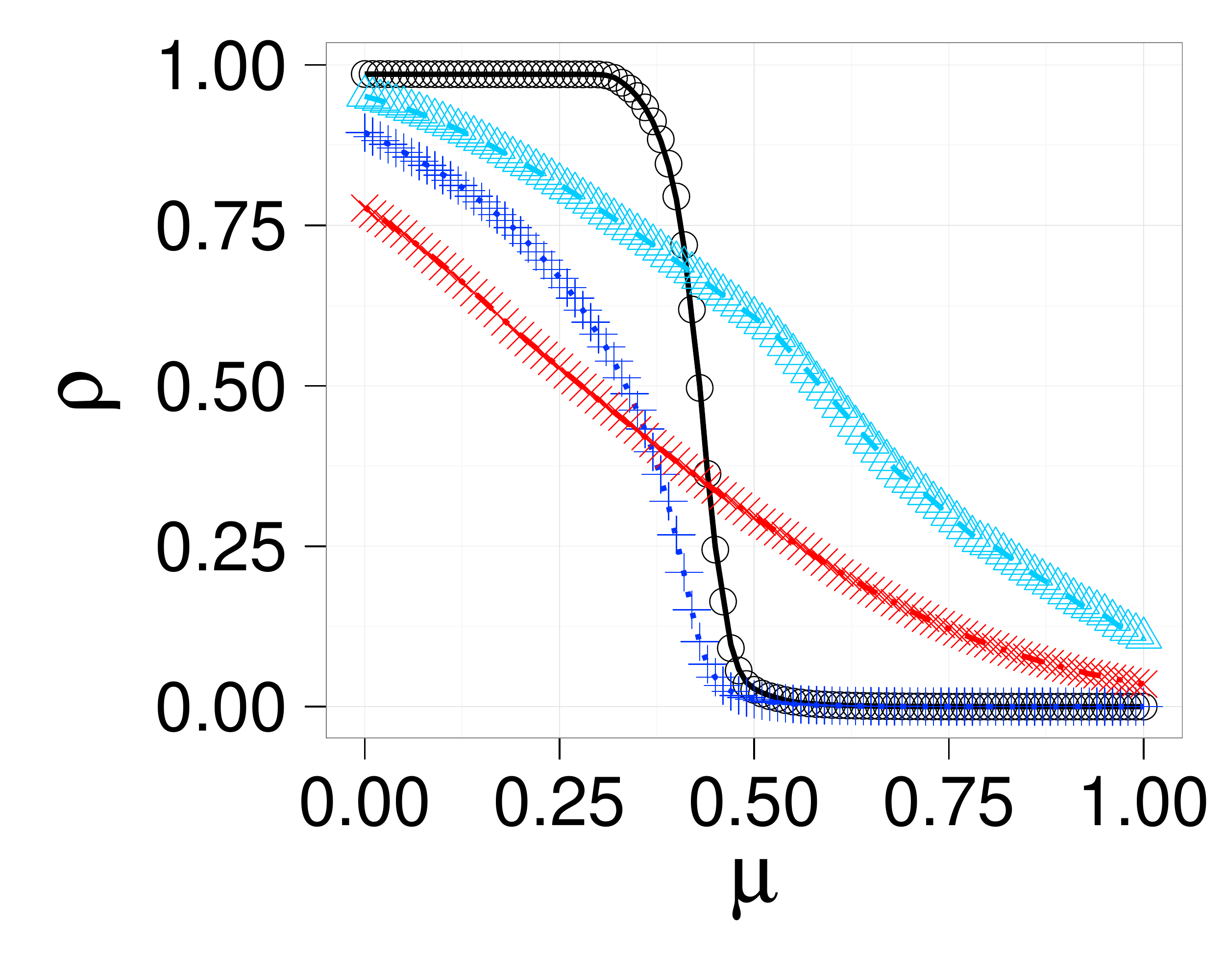}
\begin{picture}(0,0)
\put(-110,110){(a)}
\put(5,110){(b)}
   \end{picture}
 \caption{Comparison of the average final cascade size $\rho$ obtained from numerical analytic calculations and simulations, where lines represent the former and symbols in the same color correspond to the latter. Numerically, we consider $T=50$ cascade time steps. In simulations, we realize $100$ independent networks of size $N=10^5$ by the configuration model and draw independently $N$ normally distributed thresholds with mean $\mu$ and standard deviation $\sigma$ ($\Phi \sim \mathcal{N}(\mu, \sigma^2)$) that are assigned to nodes according to pf or cf strategies. (a) Pf:  Results for scale free networks  and $\sigma = 0.5$ are depicted in in light blue triangles for the ED model and in red $x$ for the DD model. Poisson random graphs and $\sigma = 0.2$ correspond to  black circles for ED and to dark blue plus signs $+$ for DD. (b) As in (a), but for cf.
}
\label{fig:SimTheoCorr}
\end{figure}
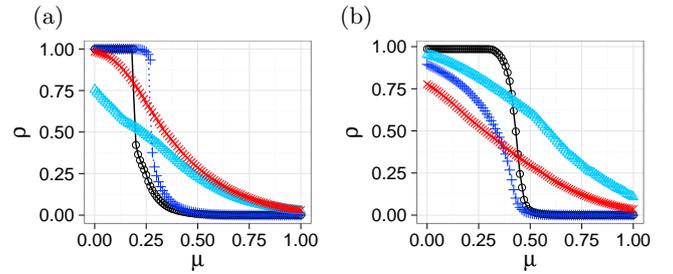

In the following, we explore different combinations of the two cascade models (ED, DD) and the two threshold allocation strategies (cf, pf).
We use a standard set-up \cite{Burkholz2015,Burkholz2017}, i.e. we calculate 
average cascade sizes on random graphs ensembles with Poisson degree distribution $p(k) \sim \lambda^k/k!$ or on scale free networks with $p(k) \sim k^{-3}$, both with average degree $z = \sum_k p(k)k = 3$.
Further, we assume normally distributed thresholds $\Phi \sim \mathcal{N}(\mu, \sigma^2)$ to test the influence of mean node robustness $\mu$ and heterogeneity $\sigma$.
Our analytic results are compared with Monte Carlo simulations (see  Fig.~\ref{fig:SimTheoCorr} for details). 
We note that our calculations based on the analytic derivations perfectly match the simulation results, even at the first order phase transition.

To discuss the results in detail, we use for comparison the case of \emph{uncorrelated} thresholds and degrees \cite{Burkholz2015}. 
In this reference set-up, the DD model usually leads to smaller average cascades $\rho$ than the ED model. 
Introducing correlations between threshold and degrees, we can confirm this finding for \emph{negative correlations}, i.e. for the cf attack scheme (the higher the degree the lower the threshold).
However, if we choose positive correlations between threshold and degrees (pf), the ED model on average leads to smaller cascades than the DD model.
Hence, positive correlations can significantly improve the robustness of the ED network. 

Secondly, we compare the robustness for different network topologies and find that, 
in comparison to Poisson random graphs, scale free networks are more robust, even for attack combinations as DD/pf and ED/cf in a region of small $\mu$.
This is interesting because so far mostly the opposite case, namely the increase of systemic risk because of the presence of hubs, has been discussed. 
The cf attack scenario reflects that hubs have a high failure risk.
However, their better risk diversification, expressed by the large number of neighbors, supports the system robustness. 

\begin{figure*}[t]
 \centering
\includegraphics[width=0.31\textwidth]{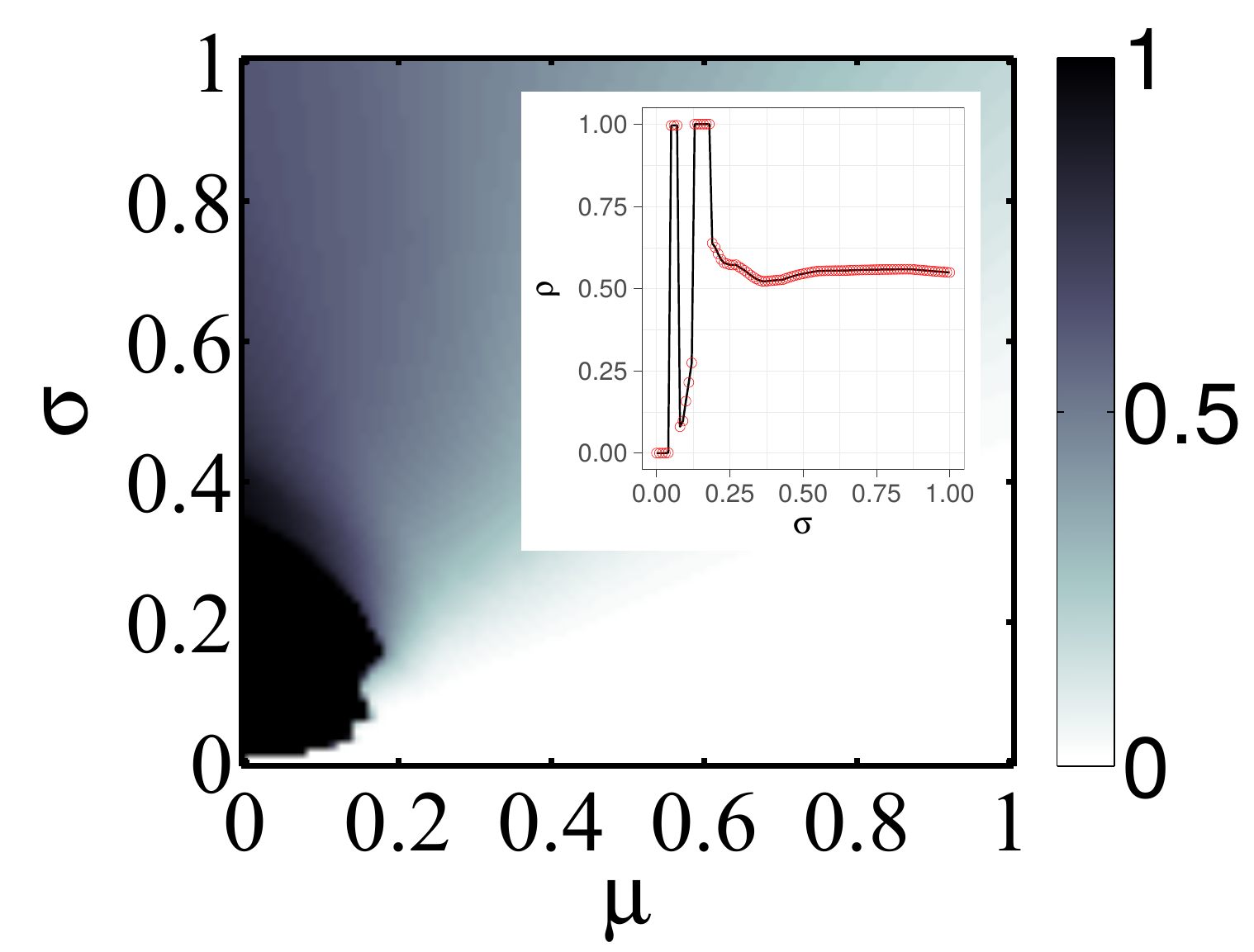}
\includegraphics[width=0.31\textwidth]{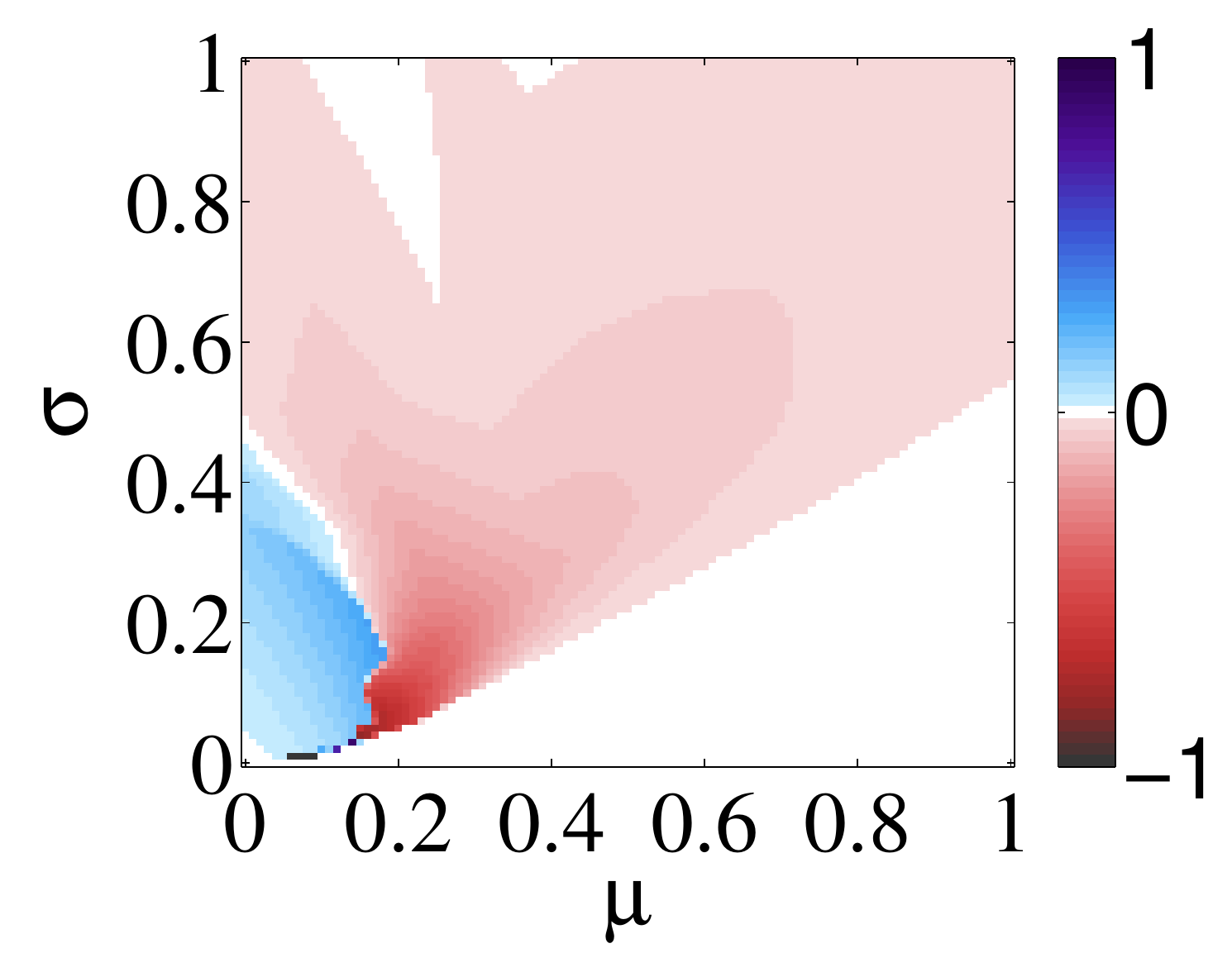}
\includegraphics[width=0.31\textwidth]{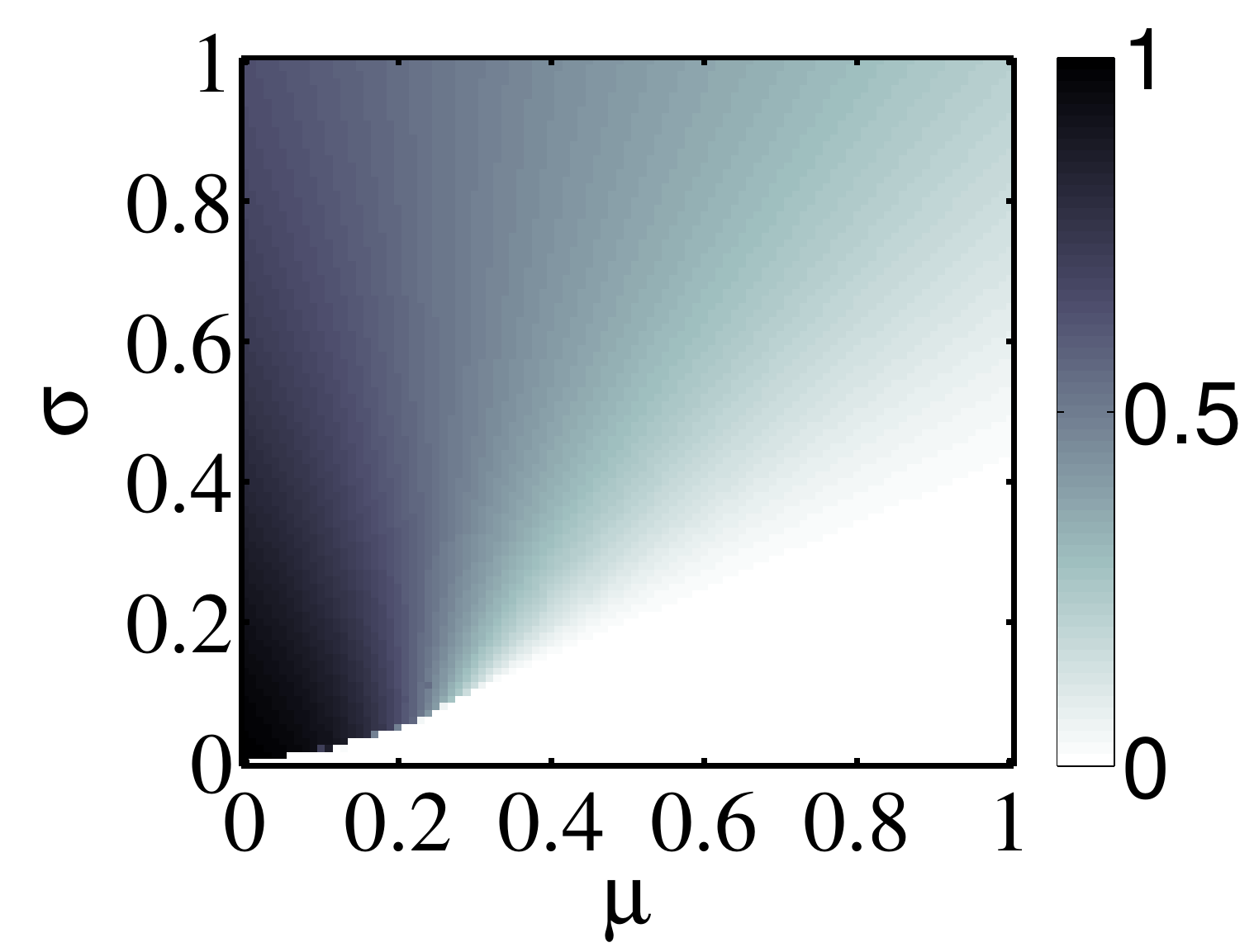}
\begin{picture}(0,0)
   \put(-470,120){(a)}
    \put(-310,120){(b)}
    \put(-150,120){(c)}
   \end{picture}
    \caption{Phase diagram for the fraction of failed nodes $\rho$ with thresholds distributed according to the order statistics obtained from a normal distribution $\mathcal{N}(\mu, \sigma^2)$ on scale free random graphs with average degree $z=3$.
    (a) ED with pf, and (c) DD with cf. The middle panel (b) shows their difference $\rho^{(a)} - \rho^{(b)}$. The inner graphic of (a) shows a snippet of (a) for $\mu = 0.17$ and varying $\sigma$, where the black curve belongs to numeric results and the red circles to simulations.}
 \label{fig:bestCorr}
\end{figure*}
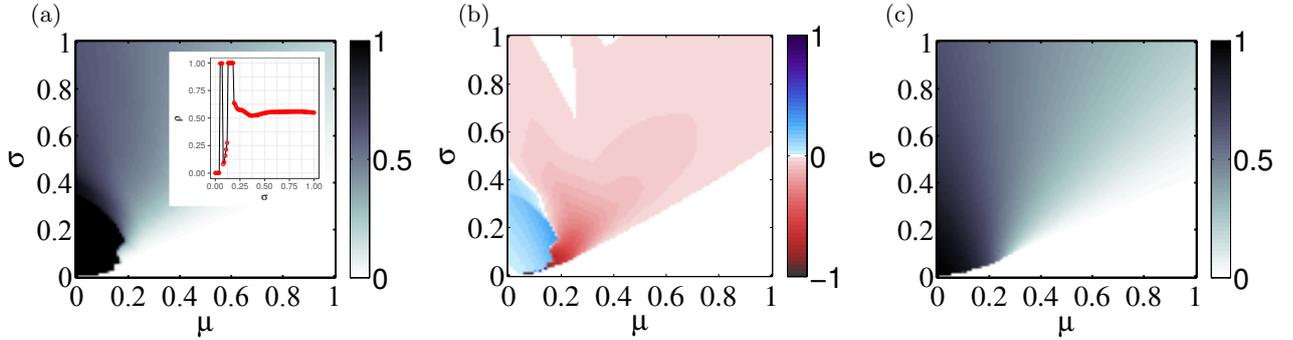

This finding is related to the \emph{robust-yet-fragile} property that scale free networks exhibit with respect to percolation \cite{Newman2010}. 
We observe a similar phenomenon for ED cascades, interestingly for \emph{both} topologies, not only for scale free networks. 
Even more, attacks can lead to larger cascades than in case of uncorrelated threshold allocations.
But this conclusion does not apply to DD models, i.e. it strongly depends on the dynamics of the cascade. 
Further, it is also sensitive to the parameters $\mu$ and $\sigma$ characterizing the threshold distribution. 
In some cases, even the combination DD/cf that is usually most robust can lead to the largest cascades. 
Still, the combination DD/cf is able to reduce the largest \emph{average cascades} for both topologies. 
Hence, it can be considered as the most promising diversification strategy for increasing systemic risk.

To shed more light on the nontrivial dependence between dynamic processes on networks (ED/DD) and attack strategies (cf/pf), we compare the phase diagram for the combination DD/cf with the combination ED/pf, which is also a risk reducing strategy. 
Fig.~\ref{fig:bestCorr} shows the results for scale free networks.
Two remarkable facts are immediately apparent. 
First, the phase transition for ED/pf in Fig.~\ref{fig:bestCorr}(a) is of irregular shape.
Usually, increasing the threshold heterogeneity $\sigma$ leads to a sudden increase of $\rho$ followed by a slow continuous decrease.
Yet, for ED/pf, $\rho$ can jump several times for small changes of $\sigma$. 
The presence of positive correlations between thresholds and degrees (pf) changes even the qualitative nature of the ED phase diagram. 
Our simulations confirm that these observations cannot be attributed to (hypothetical) numerical instabilities when calculating our approximations.
Second, none of the two studied variants, ED/pf and DD/cf, outperforms the other for all threshold parameters, as depicted in Fig.~\ref{fig:bestCorr}~(b).
While ED/pf leads to the smallest average cascades for most threshold parameters, DD/cf reduces the severity of cascades in the region of big cascades.

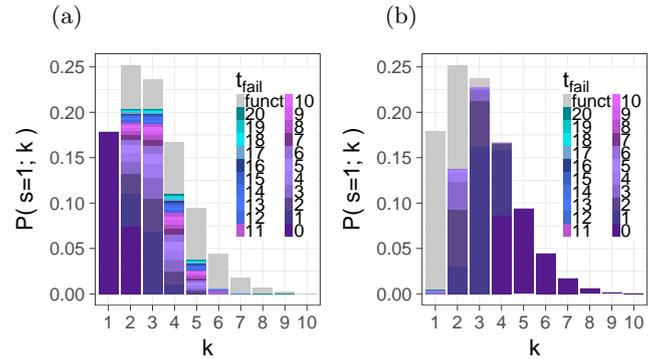
\begin{figure}[t]
 \centering
 \includegraphics[width=0.24\textwidth]{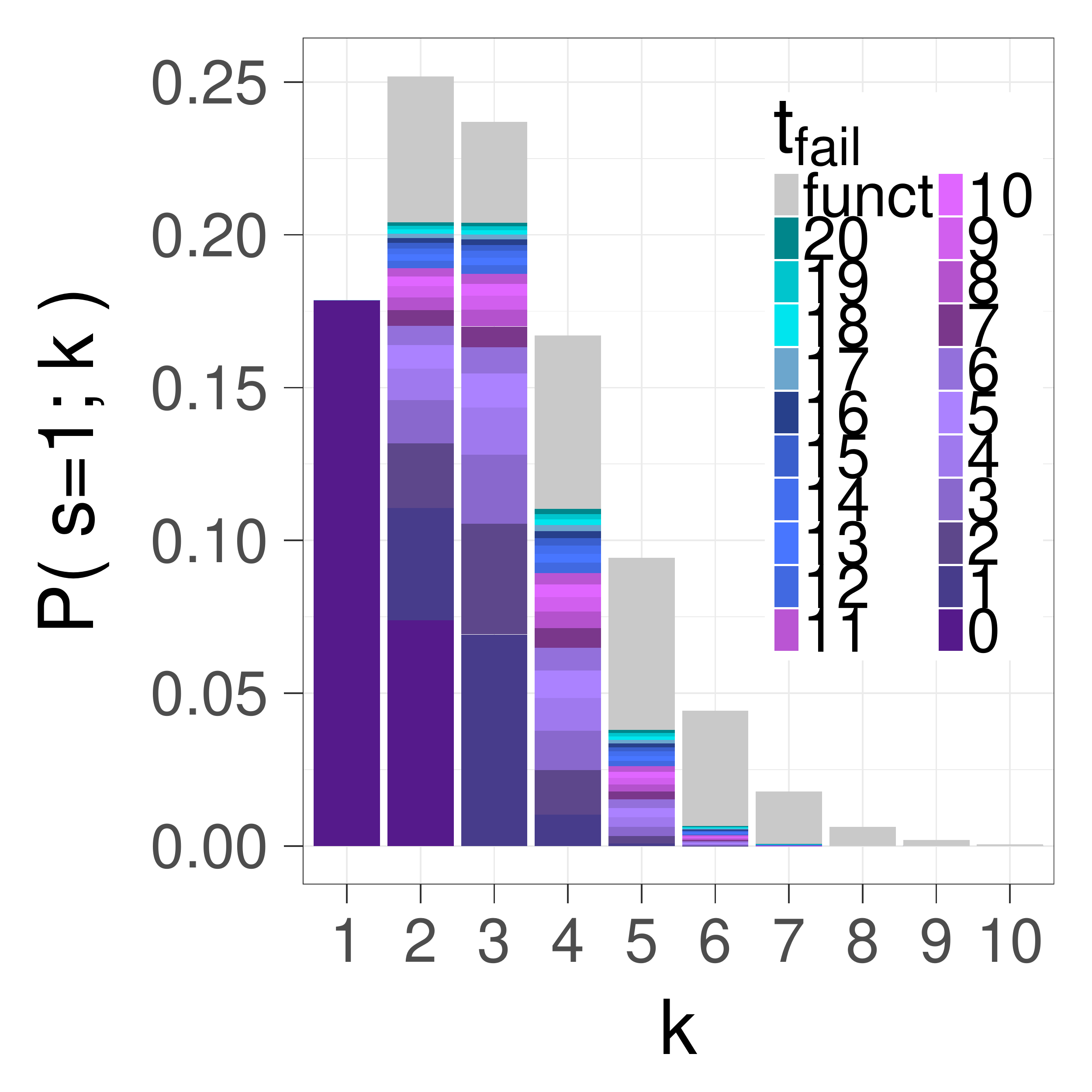}
\includegraphics[width=0.24\textwidth]{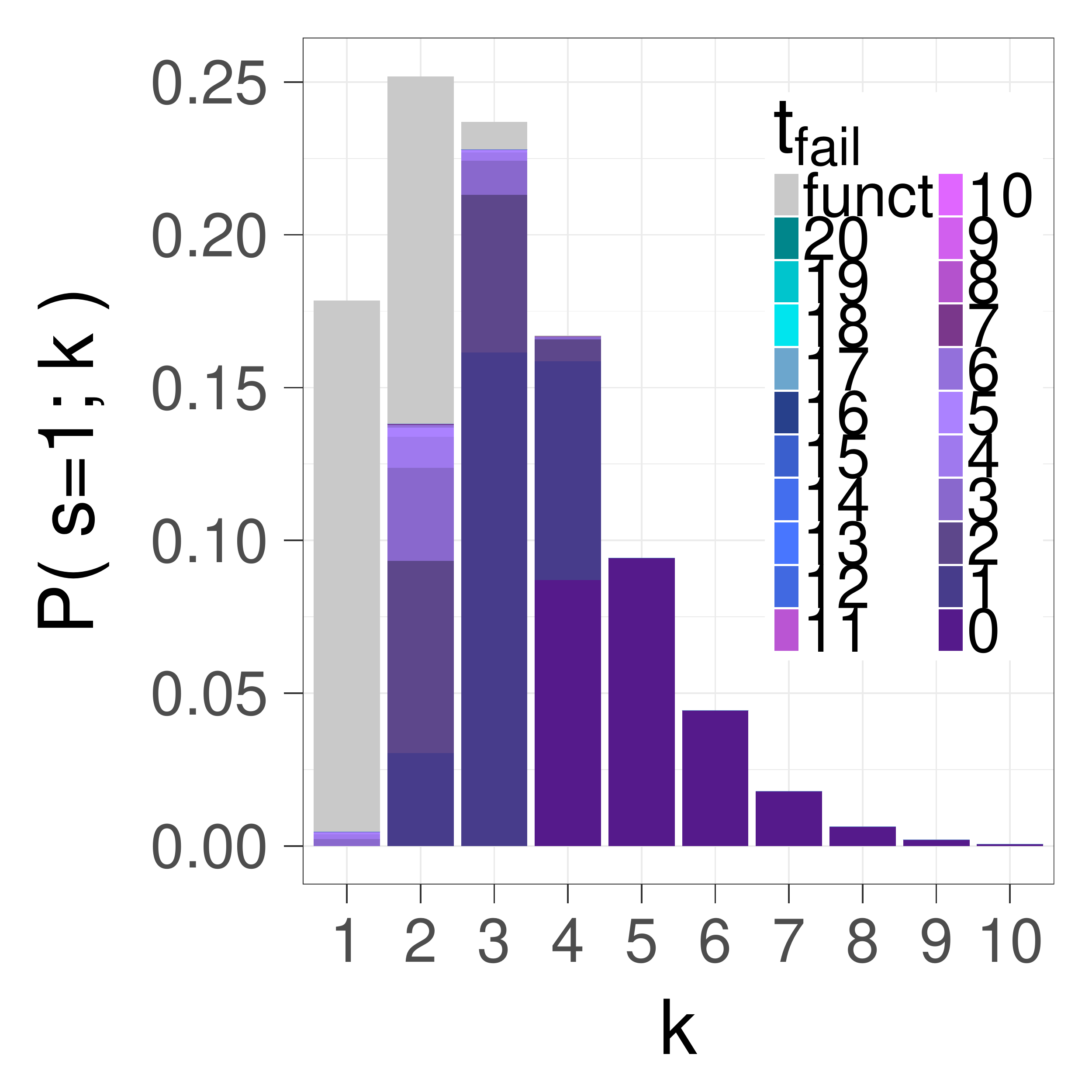}
\begin{picture}(0,0)
   \put(-105,140){(a)}
    \put(20,140){(b)}
   \end{picture}
    \caption{Fraction of failed nodes $\mathbb{P}\left(s(t)=1; k\right)$ with degree $k \leq 10$ in Poisson random graphs, which fail at the time indicated by the color of the bar. Gray corresponds to fractions of nodes that remain functional. $\Phi \sim \mathcal{N}\left(0.2, 0.3^2\right)$.
    (a) ED pf. (b) DD cf.  
  }
 \label{fig:bestCorrDistribution}
\end{figure}

Hence, in order to reduce systemic risk, system designers are confronted with two feasible options: ED/pf or DD/cf.
It is left to them to decide whether they prefer to minimize the average cascade size for most threshold parameters or to reduce the parameter space where the system breaks-down completely.
One objective excludes the other, but both cases lead to a very different outcome with respect to the surviving nodes. 
Fig.~\ref{fig:bestCorrDistribution} shows clearly that for the ED/pf combination nodes of almost every degree survive a cascade, while in the DD/cf only nodes with small degree stay functional. 
In consequence, the remaining overall connectivity is much lower for DD/cf.
This may have implications for the functionality of the overall system.
One could argue that an optimal allocation of thresholds can be solved as control problem.
Yet, our findings suggest that already small changes in the studied parameters could demand a converse control strategy. 
Thus, parameter uncertainty could hinder robust control. 
Another important aspect for dynamic controls is the different speed of failure amplification, as apparent from Fig.~\ref{fig:bestCorrDistribution} 
In the ED/pf cases, cascades tend to evolve much slower than in the DD/cf case, which leaves more time for interventions that could hinder cascades to grow larger.

In summary, we have highlighted several aspects that become relevant for cascade control based on a threshold allocation for nodes. 
Most importantly, we have shown that correlations between thresholds and degrees change considerably the occurrence of phase transitions and the overall cascade profile.
Such correlations naturally occur in engineered or self-organizing systems, hence they need to be included in any robustness analysis.
To facilitate this, we have provided an analytic description in heterogeneous mean field or local tree approximations, respectively, which becomes exact in the thermodynamic limit of infinite networks size. 
\begin{acknowledgments}
RB acknowledges financial support by the ETH48 project and Project CR12I1\_127000. FS acknowledges support by the EU-FET project MULTIPLEX 317532. 
\end{acknowledgments}

\bibliographystyle{apsrev4-1}
\bibliography{Thesis}

\end{document}